\documentclass[apjl]{emulateapj}
\slugcomment{Preprint: accepted by ApJ Letters, May 12, 2008}

\journalinfo{\scriptsize \copyright\,2008. The American Astronomical Society. All rights reserved.}

\begin{document}

\newcommand{\mic}{\ensuremath{{\rm \kern 0.2em \mu m}}}
\newcommand{\mj}{\hbox{\ensuremath{{\rm \kern 0.2em MJy \kern 0.1em sr^{-1}}}}}
\newcommand{\wms}{\hbox{\ensuremath{{\rm \kern 0.2em W \kern 0.1em m^{-2} \kern 0.1em sr^{-1}}}}}
\newcommand{\wm}{\hbox{\ensuremath{{\rm \kern 0.2em W \kern 0.1em m^{-2}}}}}
\newcommand{\wmm}{\hbox{\ensuremath{{\rm \kern 0.2em W \kern 0.1em m^{-2} \kern 0.1em \mu m^{-1}}}}}
\newcommand{\km}{\hbox{\ensuremath{{\rm \kern 0.2em km \kern 0.1em s^{-1}}}}}
\newcommand{\ccm}{\ensuremath{{\rm \kern 0.2em cm^{-3}}}}
\newcommand{\scm}{\ensuremath{{\rm \kern 0.2em cm^{-2}}}}
\newcommand{\sscm}{\ensuremath{{\rm \kern 0.2em cm^{-2}}}}
\newcommand{\icm}{\ensuremath{{\rm \kern 0.2em cm^{-1}}}}
\newcommand{\is}{\ensuremath{{\rm \kern 0.2em s^{-1}}}}
\newcommand{\htwo}{\hbox{\ensuremath{\rm H_2}}}
\newcommand{\water}{\hbox{\ensuremath{\rm H_2O}}}
\newcommand{\waterit}{\hbox{\ensuremath{H_2O}}}
\newcommand{\hii}{\hbox{\ion{H}{2}}}
\newcommand{\oiiif}{\hbox{\ion{[O}{3}]}}
\newcommand{\sif}{\hbox{\ion{[S}{1}]}}
\newcommand{\siif}{\hbox{\ion{[S}{2}]}}
\newcommand{\siiif}{\hbox{\ion{[S}{3}]}}
\newcommand{\siliif}{\hbox{\ion{[Si}{2}]}}
\newcommand{\hi}{\hbox{\ion{H}{1}}}
\newcommand{\feiif}{\hbox{\ion{[Fe}{2}]}}
\newcommand{\feiiif}{\hbox{\ion{[Fe}{3}]}}
\newcommand{\fevif}{\hbox{\ion{[Fe}{6}]}}
\newcommand{\neiif}{\hbox{\ion{[Ne}{2}]}}
\newcommand{\neiiif}{\hbox{\ion{[Ne}{3}]}}
\newcommand{\oivf}{\hbox{\ion{[O}{4}]}}
\newcommand{\clif}{\hbox{\ion{[Cl}{1}]}}
\newcommand{\hh}{\ensuremath{\rm HH\,211}}
\newcommand{\ha}{\hbox{\ensuremath{{\rm H \kern 0.1em \alpha}}}}
\newcommand{\lya}{\hbox{\ensuremath{{\rm Ly \kern 0.1em \alpha}}}}

\shorttitle{Superthermal OH in HH\,211}
\shortauthors{Tappe, Lada, Black \& Muench}

\title{Discovery of superthermal hydroxyl (OH) in the HH\,211 outflow}

\author{A.~Tappe\altaffilmark{1}, C.~J.~Lada\altaffilmark{1},  J.~H.~Black\altaffilmark{2}, A.~A.~Muench\altaffilmark{1}}
\altaffiltext{1}{Harvard-Smithsonian Center for Astrophysics, 60 Garden Street, MS-72, Cambridge, MA\,02138; {\em
atappe@cfa.harvard.edu}} \altaffiltext{2}{Onsala Space Observatory, Chalmers University of Technology, SE-439\,92
Onsala, Sweden}

\begin{abstract}
We present a 5--37\mic\ infrared spectrum obtained with the {\it Spitzer Space Telescope} toward the southeastern
lobe of the young protostellar outflow \hh. The spectrum shows an extraordinary sequence of OH emission lines
arising in highly excited rotational levels up to an energy  $E/\rm \,k\approx 28200$\,K above the ground level.
This is, to our knowledge, by far the highest rotational excitation of OH observed outside Earth. The spectrum
also contains several pure rotational transitions of \water\ ($v=0$), \htwo\ ($v=0$) S(0) to S(7), HD ($v=0$) R(3)
to R(6), and atomic fine-structure lines of \feiif, \siliif, \neiif, \sif, and \clif. The origin of the highly
excited OH emission is most likely the photodissociation of \water\ by the UV radiation generated in the terminal
outflow shock of \hh.

\end{abstract}

\keywords{ISM: Herbig-Haro objects --- ISM: individual (\hh) --- ISM: jets and outflows --- ISM: molecules  ---
shock waves}

\section{Introduction}
\label{sec;introduction}
OH and \water\ are molecules of central importance to the interstellar oxygen chemistry in many diverse
environments ranging from interstellar clouds to protoplanetary disks and comets, and they act as important shock
coolants due to their rich infrared spectra \citep[e.g.][]{holle1979,draine1983,neufe1989i,holle1989,wardl1999}.
The potential pathways leading to the formation and destruction of these molecules are now believed to be well
established. However, the relative importance of these pathways in a given astrophysical environment is generally
poorly constrained due to a lack of suitable observations and the complex interaction of formation, destruction,
and excitation mechanisms for \water\ and OH.

Both molecules are expected to be formed in abundance in hot molecular gas ($T~\raisebox{0.5mm}{\scriptsize
${\gtrsim}$}~1000$\,K) owing to a series of neutral-neutral reactions whose activation barriers are overcome at
high temperatures. {\it ISO} (Infrared Space Observatory), {\it SWAS} (Submillimeter Wave Astronomy Satellite),
and $Odin$ observations show enhanced OH and \water\ abundances in stellar outflows
\citep[e.g.][\citealt{bened2002}, \citealt{lerat2006}, \citealt{perss2007}, and \citealt{frank2007}]{giann2001}.
However, the large beam sizes prevented a detailed spatial analysis, which is essential to investigate the OH and
\water\ formation processes that lead to the increased abundances. In addition, the submillimeter observations of
\water\ are mostly confined to the $1_{10}$--$1_{01}$ transition of ortho-\water\ at $556.9$\,GHz (\hbox{$E/\rm
\,k=61$}\,K), and the wavelength coverage of the {\it ISO} Long Wavelength Spectrometer (LWS) limited observations
of OH to rotational states with $E/\rm \,k~\raisebox{0.5mm}{\scriptsize ${<}$}~1200$\,K.

In this paper, we present the first detection of rotationally excited OH at previously unobserved high excitation
levels up to $E/\rm \,k\approx28200$\,K in \hh\ with the {\it Spitzer Space Telescope} \citep{werne2004}. \hh, one
of the youngest known stellar outflows, is highly collimated, of extremely high velocity (EHV; jet speed of
100--300\km\ assuming an inclination of 5 to 10\arcdeg\ with respect to the plane of the sky, see
\citealt{oconn2005} and \citealt{lee2007}), bipolar, and
associated with a Class\,0 protostar in the young stellar cluster IC\,348. %\citep{bachi1996,froeb2005}
It has been studied in detail via excited \htwo, CO, and SiO
\citep[e.g.][]{mccau1994,gueth1999,eislo2003,oconn2005,carat2006,tafal2006,lee2007}. The unique combination of
sensitivity, wavelength coverage, and mapping capabilities with {\it Spitzer} enables us to study the spatial
structure of the \hh\ outflow bow-shock and to investigate the regions of shock-induced OH and \water\ formation.

%=========================================================================================================================================
\section{Observations and data reduction}
\label{sec;observations}
\begin{figure}[btp]
%\epsscale{1.15}
\centering
 \plotone{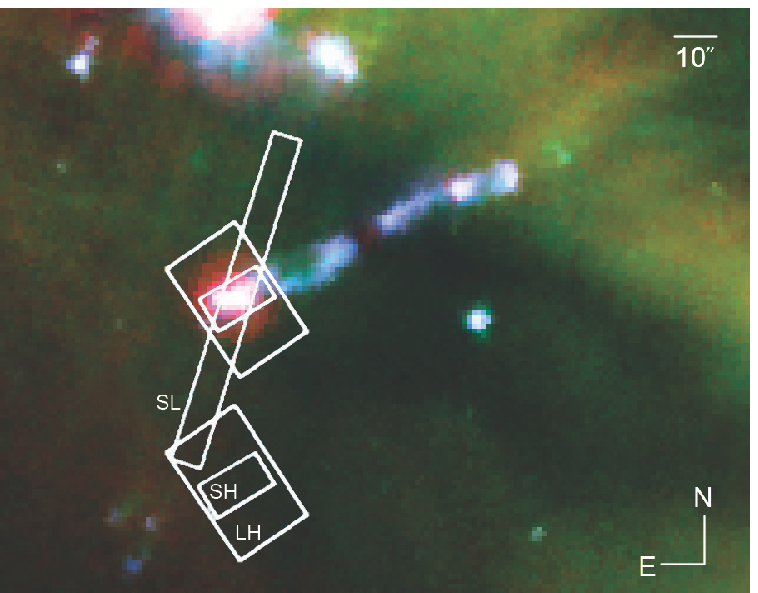} \caption{\hh: {\it Spitzer} IRAC 3--9\mic$+$MIPS 24\mic\ color composite image
 (3.6$+$4.5\mic=blue, 8.0\mic=green, and 24\mic=red). White boxes outline the IRS SL, SH, and LH map coverage including the SH$\!$/LH background
 position. The image center coordinates are $\rm3^h43^m56\fs8,\,32\arcdeg00\arcmin34\farcs4$ (J2000).}
 \label{fig:hh211}
\end{figure}

\hh\ was observed with the {\it Spitzer} Infrared Spectrograph \citep[IRS,][]{houck2004} on 2007 March~12. We
mapped the outermost region of the southeastern outflow lobe using the IRS short-low (SL2/SL1,
5.2--8.7/7.4--14.5\mic), short-high (SH, 9.9--19.6\mic), and long-high (LH, 18.7--37.2\mic) modules. The nominal
spectral resolution is $R=64$--128 as a function of wavelength for the low resolution and $R=600$ for the high
resolution settings. The total exposure times were 588\,s for each SL, 720\,s for the SH, and 600\,s for the LH
module. We also obtained IRAC \citep[Infrared Array Camera,][]{fazio2004} and MIPS \citep[Multiband Imaging
Photometer for Spitzer,][]{rieke2004} image mosaics of \hh\ from the {\it Spitzer} data archive.

We performed the IRS data reduction consisting of background subtraction, masking of bad pixels, extracting the
spectra, and generating the spectral line maps with CUBISM v1.50 \citep{smith2007}. We applied CUBISM's slit-loss
correction function, and expect the absolute flux calibration of our spectrum to be accurate within 20\%. A
calibration to the observed MIPS 24\mic\ photometry is unreliable due to the strong line emission and the more
than 2 year time difference between our IRS and the archival MIPS observations.

%=========================================================================================================================================
\section{Results}
\label{sec;results}
\begin{figure*}[btb]
%\epsscale{1.15}
\centering
 \plotone{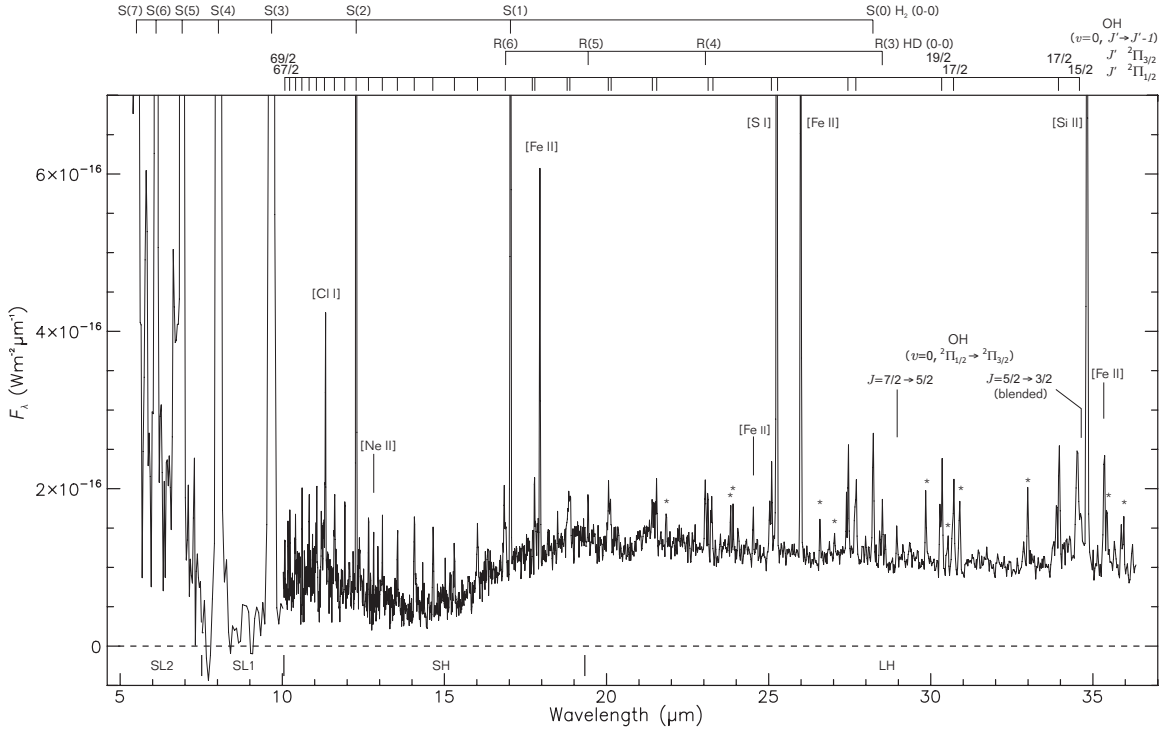} \caption{Background subtracted {\it Spitzer} IRS spectrum of \hh:
 all major detected lines are labelled, and the strongest \water\ lines are marked with an asterisk. The strongest
 lines are clipped for illustration purposes.}
 \label{fig:hh211spec}
\end{figure*}

\begin{figure}[btb]
%\epsscale{1.15}
\centering
 \plotone{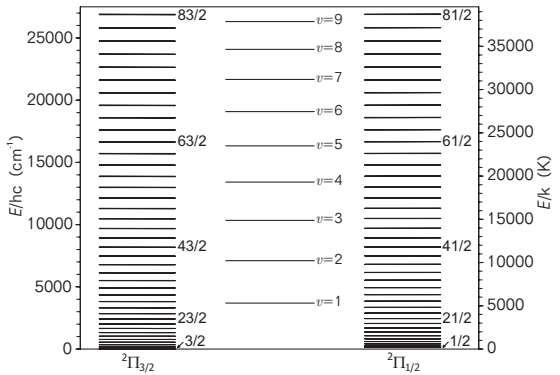} \caption{OH ($X\,^2\Pi,v\!=\!0$) rotational energy levels. The two rotational ladders result
from the spin-orbit coupling of the unpaired $2p$ electron. All rotational levels are further split into
$\Lambda$-type doublets with separations between 0.1 and 35.0\icm, which are not visible on the scale of this plot
\citep[data from][]{colin2002}.}
 \label{fig:OHenergy}
\end{figure}

The spectrum of \hh\ displayed in Figure~\ref{fig:hh211spec} reveals a sequence of highly excited OH ($v=0$, $J'
\rightarrow J'-1$) pure rotational transitions arising in the $^2\Pi_{3/2}$ and $^2\Pi_{1/2}$ rotational ladders
from \hbox{$J'=15/2$} to 69/2, which result in closely spaced doublets that become partly resolvable only at
wavelengths greater than 20\mic\ in our spectra. An additional splitting due to $\Lambda$-doubling is only barely
noticeable for the OH lines at the longest wavelengths in our spectra \citep[see][for a detailed treatment of OH
spectroscopy]{herzb1971}. The OH energy level diagram in Figure~\ref{fig:OHenergy} shows that the highest excited
OH transition has an upper state energy of $E/\rm \,hc=19607$\icm\ ($E/\rm \,k\approx28200$\,K) above the ground
level. Note that the population of the OH energy levels probably extends to even higher energies, but the
corresponding rotational transitions fall in the wavelength range of the low resolution IRS modules below 10\mic.
Due to the lower resolution, the SL modules are not sensitive enough to detect the faint, narrow emission lines.

In addition, there are OH cross-ladder transitions with $\Delta J=0,\,+1$, but all such transitions that fall in
the 5-37\mic\ region have transition probabilities orders of magnitude smaller than the intra-ladder transitions
described above. Nevertheless, we detect the two lowest excited OH ($v=0$, $ ^2\Pi_{1/2} \rightarrow
{^{2}\Pi}_{3/2}$, $J' \rightarrow J'-1$) cross-ladder transitions with upper levels $J'=7/2$ and $5/2$ ($E/\rm
\,k\approx620$ and 420\,K) at 28.94 and 34.62\mic\ (see Fig.~\ref{fig:hh211spec}), respectively, indicating a
substantial population in the lower OH $J$-levels. We did not unambiguously detect any pure rotational OH
transitions from excited vibrational levels.

The low resolution portion of the spectrum is dominated by the strong, pure rotational \htwo\ (0--0) S(3)--S(7)
transitions and the much weaker \feiif\ 5.34\mic\ line. In addition to the previously mentioned OH lines, the high
resolution portion longward of 10\mic\ shows \htwo\ (0--0) S(0)--S(2) including (1--1) S(3) at 10.18\mic, HD
(0--0) R(3)--R(6), numerous \water\ ($v=0$) pure rotational lines, and forbidden atomic fine-structure lines from
\feiif, \siliif, \neiif, \sif, and \clif. The spectrum also has noticeable continuum emission longward of 15\mic,
which can be fitted by thermal dust emission at a temperature \hbox{$\sim85$\,K}, represented by a modified
blackbody with the dust emissivities of \citet{weing2001}. Two additional components at about 170\,K and 30\,K are
needed to fit the complete continuum including the 8--15\mic\ region and the MIPS 70\mic\ flux of
$F_{\lambda}=1.1\times10^{-16}$\wmm. There is also a strongly rising continuum shortward of 8\mic. The peculiar
NIR continuum of \hh\ was previously noted by \citet{eislo2003} and confirmed by \citet{oconn2005}, who interpret
it as scattered light from the protostar escaping along the low-density jet cavity.

%=========================================================================================================================================
\section{Discussion}
\label{sec;discussion}

\subsection{The origin of high-$J$ OH emission}
\label{sec:precsursor}
In $\S\,\ref{sec;results}$, we reported the detection of rotationally excited OH at previously unobserved high
excitation levels in \hh. As a consequence of the large OH dipole moment, the observed intra-ladder pure
rotational transitions have large Einstein $A$-coefficients of the order of 10 to 400\is. For comparison, our
observed pure rotational \water\ transitions have $A\sim1$--$20$\is, the \feiif\ lines have $A$-values of a few
$10^{-3}$\is, and the \htwo\ S(0)--S(7) pure rotational lines have $A$-values between $3\times10^{-7}$ and
$3\times10^{-11}$\is. Large $A$-values make collisional excitation of high energy levels very ineffective at lower
densities, i.e.~the corresponding critical densities are usually much larger than the gas density. The observed OH
lines in \hh\ demonstrate that energy levels up to at least $E/\rm \,k\approx28200$\,K are well populated despite
the large OH $A$-values. This fact, coupled with the observed distribution of OH line intensities, strongly
suggests a selective, non-thermal origin of the high-$J$ OH excitation. For comparison, a non-LTE analysis of the
\feiif\ lines in our spectrum suggests a gas temperature less than a few 1000\,K. Furthermore, all of our observed
pure rotational \water\ lines originate from energy levels with $E/\rm \,k~\raisebox{0.5mm}{\scriptsize
${<}$}~2400$\,K even though there are many pure rotational \water\ transitions with $A\sim50$--$100$\is\ from
energy levels between 3000 and 8000\,K in the 10--38\mic\ wavelength range.

Theoretical and experimental laboratory studies show that OH molecules produced via photodissociation of \water\
at photon energies larger than about 9\,eV ($\lambda < 140$\,nm) are mostly in the ground electronic and
vibrational state but with a high rotational excitation favoring $J=70/2$ to $90/2$ with $E/\rm
\,k\,\raisebox{0.5mm}{\scriptsize ${\gtrsim}$}\,30000$\,K \citep{morda1994,harre2000,haric2000}. This is because
the absorption of UV radiation in the \water\ ($\tilde{B}$--$\tilde {X}$) band produces OH in an excited
electronic state, $A\ ^2\Sigma ^ +$, which, in turn, efficiently yields highly excited ground-state OH ($X\
^2\Pi$) molecules via a nonadiabatic crossing between intersecting potential energy surfaces
\citep[see][]{harre2000,haric2000}. This process leads to a highly selective formation of high-$J$ OH in the
ground electronic and vibrational state, followed by a radiative cascade down the rotational ladders depicted in
Fig.~\ref{fig:OHenergy}. This is consistent with our observed distribution of OH line intensities and our
non-detection of OH pure rotational transitions from excited vibrational states. We will present a detailed
non-LTE excitation model of OH, \water, and \htwo\ together with an analysis of the atomic fine-structure
transitions in a subsequent paper.

\begin{figure*}[btb]
%\epsscale{1.15}
\centering
 \plotone{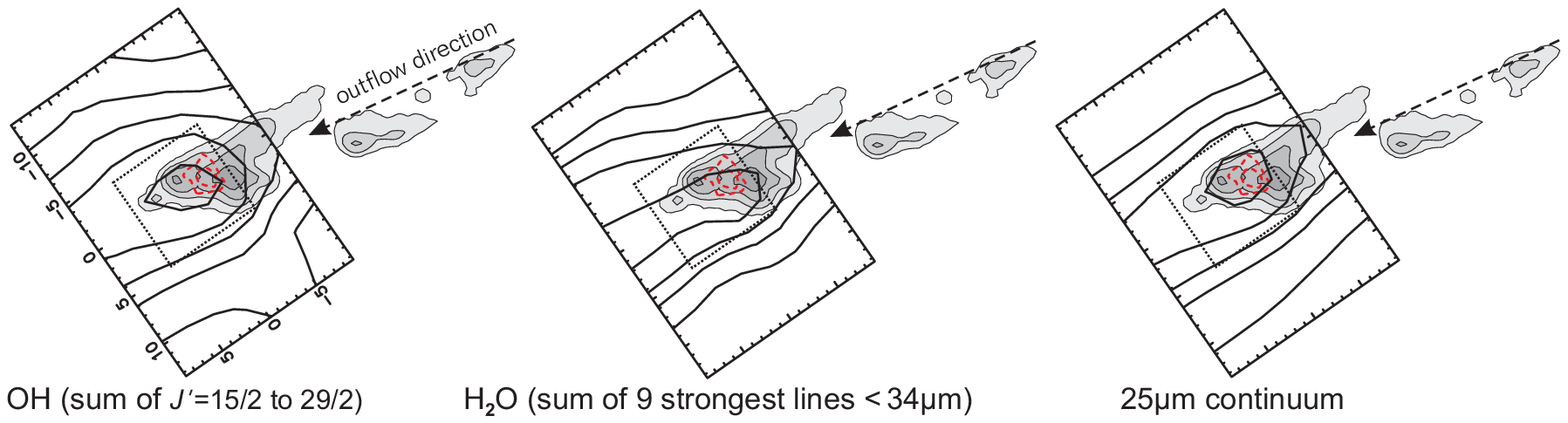} \caption{{\it Spitzer} IRS LH emission line maps of \hh\ (solid lines) overlaid on the \htwo(1--0) S(1)
2.12\mic\ \citep[shaded contours;][]{eislo2003} and optical \ha\ emission \citep[red, dashed
contours;][]{walaw2006}. Coordinates are in arcseconds offset from the (0,0) center. The maps are interpolated to
half the original LH pixel size of 4.5\arcsec, and the positional accuracy is $\sim\!1\arcsec$. The dotted box
marks the extracted region of the LH spectrum in Fig.~\ref{fig:hh211spec}. The linearly spaced max/min levels for
the OH and \water\ emission contours going outwards from the center are $5.2/1.2\times10^{-8}$ and
$1.0/0.2\times10^{-8}$\wms, and 13/2\mj\ for the 25\mic\ continuum.}
 \label{fig:maps}
\end{figure*}

Further evidence for the role of \water\ photodissociation as a source of high-$J$ OH can be gained from spatial
mapping of the emission lines. Our {\it Spitzer} IRS maps cover the southeastern terminal shock of \hh, which
exhibits a typical bow-shock geometry in \htwo\ and CO emission \citep{gueth1999,eislo2003}. Figure~\ref{fig:maps}
shows the spatial distribution of rotationally excited OH and \water\ together with the 25\mic\ dust continuum
surface brightness. The highly localized optical \ha\ emission \citep{walaw2006} is practically coincident with
the 25\mic\ surface brightness maximum. Their emission maxima are offset by about 1\arcsec, which is within the
positional map accuracy (see Fig.~\ref{fig:maps}). This region is located at the bow-shock apex, where the shock
velocity is expected to be the highest and presumably most of the shock generated FUV radiation originates. The
emission maximum of rotationally excited OH (Fig.~\ref{fig:maps}, left map) also coincides with the bow shock
apex. Although not shown in Figure~\ref{fig:maps} due to the limited spatial coverage, the emission maximum of the
OH ($J'~\raisebox{0.5mm}{\scriptsize ${>}$}~29/2$) rotational lines in the SH wavelength range occurs at the same
location. However, unlike \ha, the continuum, OH, and \water\ emission clearly extend well beyond the bow shock.
Using the 25\mic\ continuum surface brightness as a surrogate tracer of the UV radiation field, we clearly note
the presence of a radiative precursor upstream of the shock. Note that although collisional grain heating due to
inelastic gas-grain collisions may contribute to the dust heating in the bow shock, it is probably unimportant
outside and ahead of the shock.

\subsubsection{Formation of OH and \waterit}
Our emission line maps show clear evidence of OH and \water\ emission in the precursor region (see
Fig.~\ref{fig:maps}). It is unlikely that the chemical formation route via $\rm \htwo + O \rightarrow OH + H$ and
$\rm \htwo + OH \rightarrow \water + H$ \citep[e.g.][]{holle1979,neufe1989i} is the dominant source of OH and
\water\ in the UV precursor upstream of the shock. These reactions have significant Arrhenius activation energies
($E_{\rm A}/\rm \,k$ about 3150 and 1736\,K, respectively) and only proceed efficiently in dense, hot gas. The
absence of rovibrational \htwo\ emission in the preshock region confirms the absence of hot, shocked gas needed to
directly form OH and \water\ through gas chemistry.

Instead, we propose that UV induced photodesorption of water ice from grain mantles and photodissociation of
\water\ either in the gas phase or directly in the grain ice mantles \citep[see][]{ander2006} are the primary
sources of OH and \water\ in the precursor. We evaluated the \water\ photodesorption rate using the \lya\
dominated FUV emission generated by a shock with a speed of \hbox{$v_{\rm s}=40$\km} and a preshock density of
\hbox{$n_{\rm 0}=10^3$--$10^4$\ccm} (atomic shock models by J.~Raymond, private communication). Such a shock
approximately reproduces our observed \neiif\,12.8, \sif\,25.2, \feiif\,26.0, and \siliif\ 34.8\mic\ emission line
intensities, which are good indicators for $v_{\rm s}$ and $n_{\rm 0}$ \citep[see e.g.~Fig.~7 in][]{holle1989}.
Our estimate shows that grain ice mantles can be efficiently removed via photodesorption on timescales of the
order of \hbox{$10^3$\,yr} assuming the experimentally measured \lya\ photodesorption yield for water ice of
3--$5\times10^{-3}$\,molecules per photon and a \lya\ photodesorption cross section
\hbox{$\sim8\times10^{-18}$\sscm} per individual water ice molecule \citep{westl1995ii,westl1995i}. This is
consistent with the estimate of \citet{holle1979}, who argued that a shock with $v_{\rm
s}\,\raisebox{0.5mm}{\scriptsize ${\gtrsim}$}\,40$\km\ would lead to complete upstream photodesorption of grain
ice mantles assuming a photodesorption yield of $5\times10^{-3}$. At the high visual extinction toward the
terminal shock of the \hh\ southeastern lobe, $A_{\rm V}\approx12\pm3$\,mag \citep{oconn2005,carat2006}, most of
the gas phase water would otherwise be expected to reside frozen onto grains \citep[cf.][]{melni2005}.

It is interesting to compare our observation of \hh\ with a recent {\it Spitzer} IRS detection of strong water
emission toward the protostellar object IRAS~4B in NGC\,1333 by \citet{watso2007}. These authors attribute the
origin of this emission to the very dense, warm molecular layer of a protoplanetary disk. In this case, the main
source of water is probably thermal sublimation of grain ice mantles in the \hbox{$\sim170$\,K} warm surface layer
as suggested by \citet{watso2007}. Sublimation is insignificant at grain temperatures below about 100\,K (cf.~the
derived dust temperature of $\sim85$\,K in \hh\ from our dust continuum fit, see $\S\,\ref{sec;results}$), but it
becomes rapid above 120\,K \citep[see][Table 2]{frase2001}. In addition to the strong \water\ emission, we have
subsequently identified OH emission limited to transitions with $J'~\raisebox{0.5mm}{\scriptsize ${<}$}~21/2$
($E/\rm \,k~\raisebox{0.5mm}{\scriptsize ${<}$}~3500$\,K) in the IRS LH spectrum of IRAS~4B by reanalyzing the
{\it Spitzer} archival spectra originally published by \citet{watso2007}. The comparatively low excitation of OH
indicated by the absence of lines with \hbox{$J'=23/2$} to $27/2$ and the weakness of the OH emission relative to
\water\ in IRAS~4B are consistent with the proposed sublimation origin of water and suggest that FUV
photodissociation of \water\ plays only a minor role compared to \hh.

%----------------------------------------------------------------------------------------------------------------------------------------
\acknowledgements{This work is based on observations made with the Spitzer Space Telescope, which is operated by
the Jet Propulsion Laboratory, California Institute of Technology under a contract with NASA. Support for this
work was provided by NASA, by a {\it Spitzer} GO grant (JPL \#1288815), and by the Swedish Research Council. We
thank J.~Eisl\"{o}ffel, D.~Froebrich, and K.-W.~Hodapp for providing the near-IR \htwo\ data, J.~Walawender for the
optical \ha\ data used in Fig.~\ref{fig:maps}, and J.~Raymond for providing shock model results and useful
discussions. Furthermore, we thank D.~Neufeld for informative discussions, in particular, for calling our
attention to the laboratory experiments of \water\ photodissociation and for suggesting the possible relevance of
the OH excitation mechanism through \water\ photodissociation for HH211. }

%=========================================================================================================================================
%=========================================================================================================================================

\end{document}